\documentclass[amsfonts, amssymb, amsmath, aip, reprint, showkeys, nofootinbib, twoside, superscriptaddress]{revtex4-2}
\usepackage[english]{babel}
\usepackage[utf8]{inputenc}
\usepackage{dcolumn}
\usepackage{graphicx}
\usepackage{bm}
\usepackage[pdftex, pdftitle={Article}, pdfauthor={Author}]{hyperref} 
\begin{document}
\title{Ultrasound differential phase contrast using backscattering and the \\memory effect}

\author{Timothy D. Weber}
\affiliation{Dept. of Biomedical Engineering, Boston University, Boston MA 02215, USA}
\author{Nikunj Khetan}
\affiliation{Dept. of Mechanical Engineering, Boston University, Boston MA 02215, USA}
\author{Ruohui Yang}
\affiliation{Dept. of Biomedical Engineering, Boston University, Boston MA 02215, USA}
\author{Jerome Mertz}
 \email[Corresponding author email: ]{jmertz@bu.edu}
\affiliation{Dept. of Biomedical Engineering, Boston University, Boston MA 02215, USA}


\begin{abstract}

We describe a simple and fast technique to perform ultrasound differential phase contrast (DPC) imaging in arbitrarily thick scattering media. Though configured in a reflection geometry, DPC is based on transmission imaging and is a direct analogue of optical differential interference contrast (DIC). DPC exploits the memory effect and works in combination with standard pulse-echo imaging, with no additional hardware or data requirements, enabling complementary phase contrast (in the transverse direction) without any need for intensive numerical computation. We experimentally demonstrate the principle of DPC using tissue phantoms with calibrated speed-of-sound inclusions. 

\end{abstract}


\maketitle

Ultrasound imaging is generally based on pulse-echo sonography, which excels at revealing reflecting interfaces and point-like structures in a thick sample. However, the quality of ultrasound images becomes degraded when aberrations are present in the sample, since these cause acoustic phase distortions or speed-of-sound (SoS) variations that undermine the accurate reconstruction of beamformed images \cite{anderson2000}. Longstanding efforts have gone into developing techniques to restore image quality by correcting for the effects of aberrations \cite{flax1988,nock1989,liu1994}. It has also been recognized that the aberrations themselves can be of interest, for example in helping identify soft-tissue pathologies \cite{lin1987,li2009} . 

But aberrations are difficult to directly image with ultrasound since they generally do not cause acoustic reflections. Initial attempts at imaging aberrations were based on transmission geometries using multiple receivers \cite{glozman2010,duric2013} or a passive backreflector \cite{nebeker2012}. Much more practical are techniques that use conventional pulse-echo sonography with a single ultrasound probe, since these enable imaging in arbitrarily thick samples \cite{robinson1991}. For example,  2D phase maps can be obtained from a 1D array, by inferring the local SoS based on the iterative application of a local focusing criterion \cite{imbault2017,lambert2020}. Alternatively, such maps can be obtained by a method of computed ultrasound tomography based on the numerical inversion of a forward model \cite{krucker2004,stahli2020}. These approaches are highly promising, though computationally intensive.       

\begin{figure}[tbp]
	\includegraphics[width =3.2in]{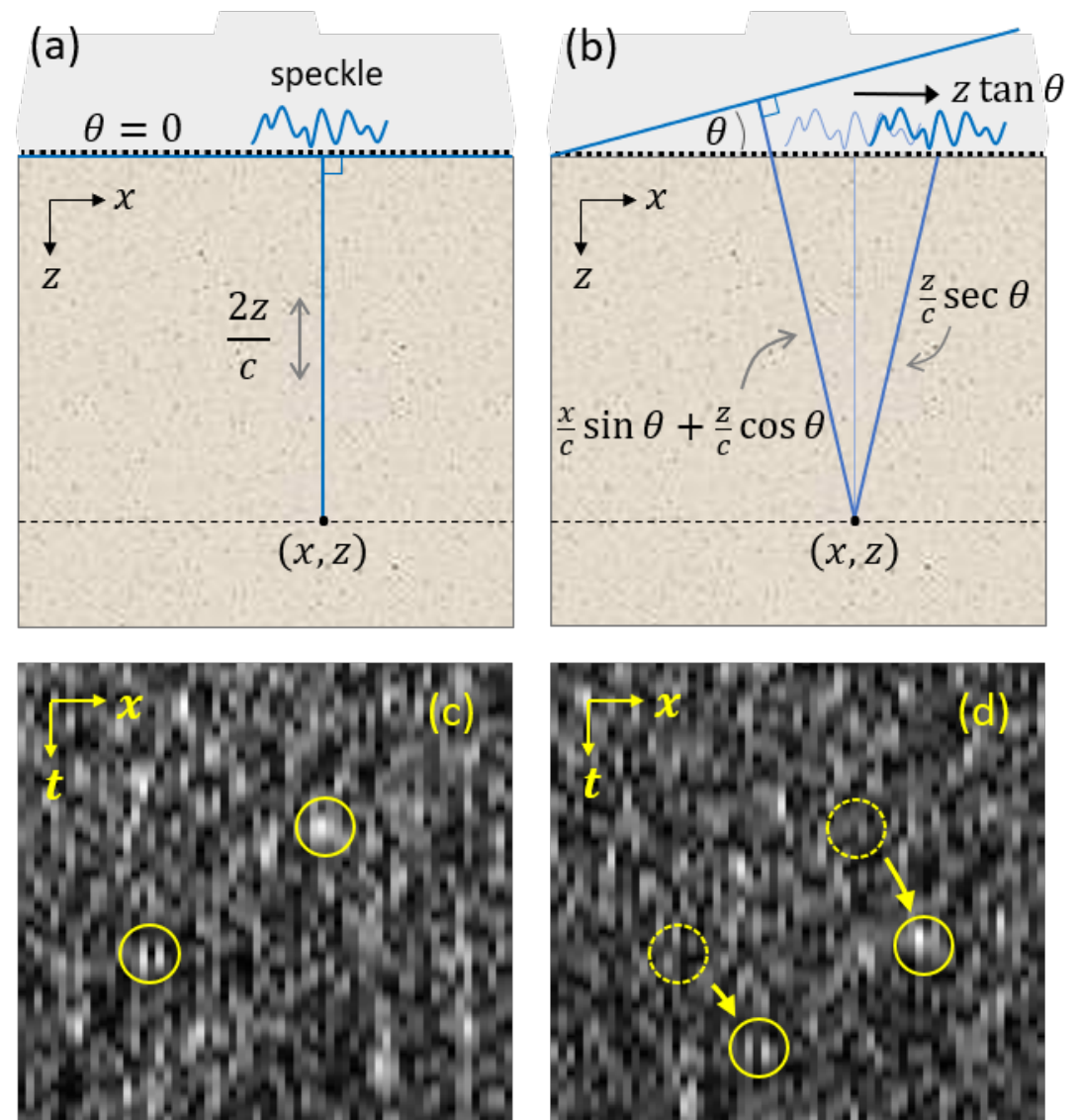}
	\caption{Demonstration of memory effect when scattering sample is insonified by US linear array (gray). (a) Transmit plane wave normal to surface is backscattered, producing a received speckle echo. Blue line shows round-trip acoustic time delay associated with arbitrary sample point $(x,z)$. (b) Speckle is translated by memory effect when transmit pulse is tilted. (c,d) Example speckle trajectories in $(x,t)$ space.}      
	\label{fig:schematic}
\end{figure}

We present here a technique to perform phase-contrast imaging directly and in real time, without the need for a forward model or intensive computation. Our technique is a direct analogue of optical differential interference contrast (DIC) microscopy, here applied to ultrasound. The principle of DIC is well known \cite{pluta1994}. In essence, a DIC microscope is a transmission-based lateral shear interferometer, which, remarkably, can perform phase contrast imaging with completely incoherent illumination. This is achieved by using a beamsplitter (a Normarski prism) to split the incoherent illumination into two identical copies that are laterally sheared relative to one another. Even though the illumination is incoherent, it remains mutually coherent with its twin copy exactly at the shear separation distance. The two copies co-propagate and transilluminate a sample, whereupon they can accumulate phase differences caused by sample structure. Finally, the copies are recombined with a second beamsplitter (a matched Nomarski prism), and their resulting interference is imaged onto a camera with a high resolution microscope, yielding an image of the phase gradients in the sample along the shear axis.

In our case, incoherent illumination (or rather insonification) is obtained by launching a plane-wave transmit pulse into a thick sample, and relying on random backscattering from the sample to render the return pulse incoherent \cite{mallart1991}. To obtain twin copies of this backscattered field that are laterally sheared relative to one another, we rely on a phenomenon called the memory effect \cite{Feng1988,berkovitz1989,freund1990}, whereupon by simply steering the tilt angle of the transmit plane wave, the backscattered field becomes laterally displaced while remaining otherwise largely unchanged (over a limited range).

An experimental demonstration of the memory effect is shown in Fig. 1. We use a Verasonics Vantage 256 equipped with a GE9L-D linear array probe (192 elements, 0.23mm pitch, 5.3 MHz center frequency, $4\times$ sampling frequency) serving as both transmitter and receiver. When a transmit plane-wave pulse is launched normal to the sample surface (CIRS Model 049 phantom -- Fig. 1a), the resulting backscattered echo $RF(x,t)$ is incoherent, with an intensity that takes on the appearance of speckle (Fig. 1c -- $x$ is the transducer coordinate and $t$ is time). Consider now the effect of electronically tilting the transmit pulse a small angle $\theta$ (Fig. 1b). Because of the memory effect, the received speckle pattern becomes shifted in $x$, but also becomes delayed in $t$ because of the additional pathlengths involved from transmit to scatter and back to receive. The memory effect stipulates that the trajectory of a speckle grain follows a mirror-reflection law \cite{berkovitz1989,freund1990}, schematically depicted in Fig. 1b. That is, a speckle grain received at coordinate $(x_0,t_0)$ when $\theta=0$ is received at coordinate $(x_{\theta},t_{\theta})$ when the transmit angle is tilted, where, from geometrical considerations

\begin{align}
		x_{\theta} = & x_0 + \frac{ct_0}{2} \tan \theta  \label{DPC1} \\ 
		t_{\theta}=& \frac{t_0}{2} (\cos \theta + \sec \theta) + \frac{x_0}{c} \sin \theta  \label{DPC2}
\end{align}

\noindent ($c$ is the speed of sound in the medium, here 1540 m/s). It should be noted that the memory effect is not exact, and that the speckle patterns remain correlated over only a limited tilt range that scales inversely with the acoustic transport length \cite{freund1990}. At first glance it might appear that this range should be small because this length is long. However, the speckle patterns here are time-resolved (or time-gated), meaning that the effective sample interaction thickness for any given speckle pattern is only a small fraction of a millimeter, significantly extending the correlation range \cite{Kadobianskyi2018}. We found that the above relations remained valid even for tilt angles as large as $\pm 12$ degrees (the maximum range we explored).

We now consider how backscattering and the memory effect enable us to perform an ultrasound analogue of optical DIC.
As noted above, DIC requires sheared copies of a same incoherent field.
In our case the incoherent field is produced by backscattering, and twin copies of this field are sheared by the memory effect (Fig. 2a). Our technique differs from optical DIC in that the shearing need not be instantaneous, but instead occurs sequentially. Moreover, the phase differences accumulated by the fields need not be measured by interference, which converts phase differences into intensity variations that can be recorded by a camera. Instead, the phases of the fields are measured directly by the transducer, and their differences evaluated numerically a posteriori. Accordingly, because our technique does not require interference, we refer to it as ultrasound differential phase contrast (US DPC), to distinguish it from optical DIC. 

\begin{figure}[tbp]
	\includegraphics[width=3.2in]{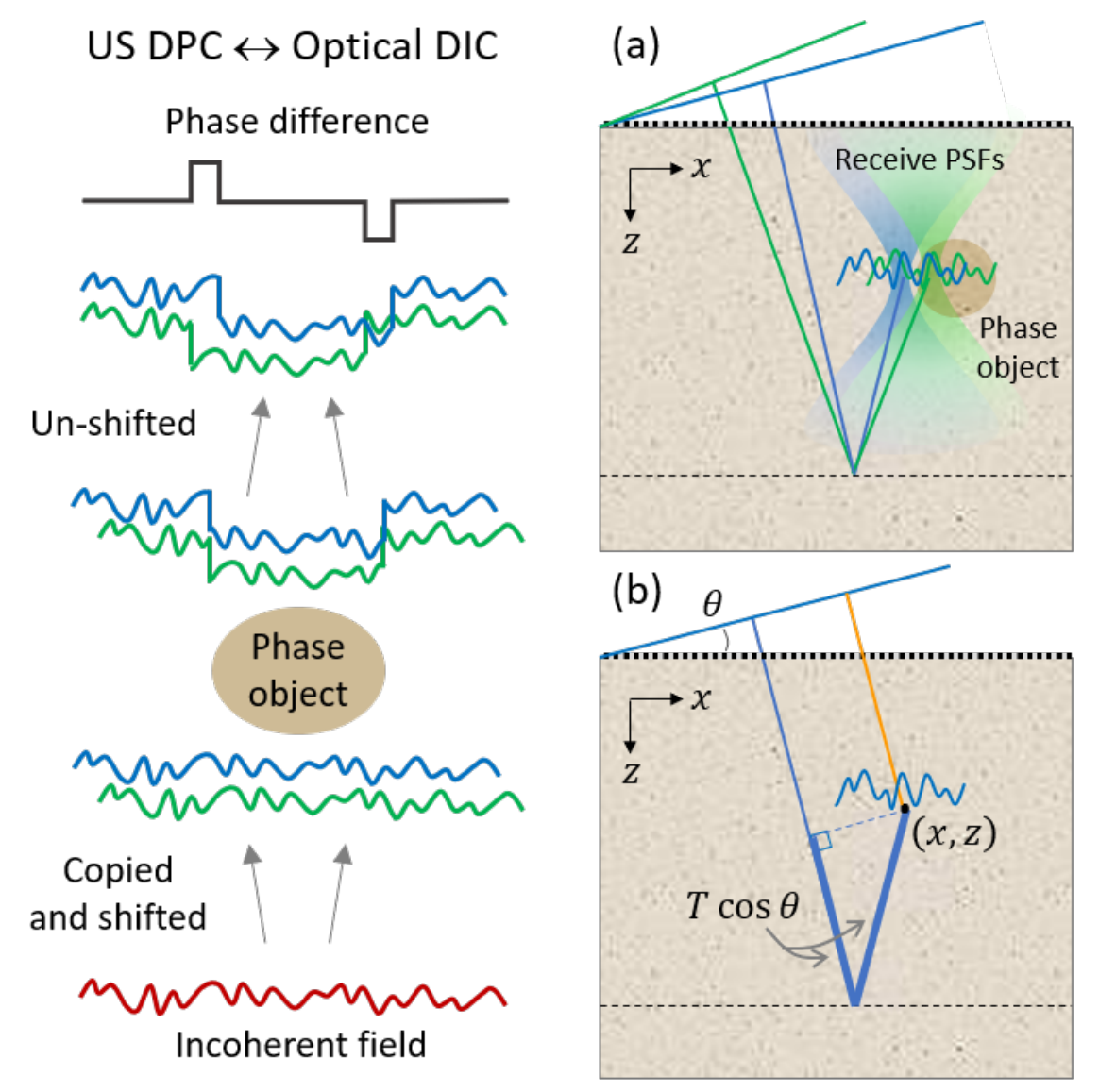}
	\caption{Analogy between US DPC and optical DIC. (a) Two transmit tilts produce sheared copies of speckle that are imaged by the receive point-spread-functions (PSFs) obtained by standard beamforming. (b) Adding a time delay $T \cos \theta$ (thick blue segments) to the receive RF ensures that an arbitrary point $(x,z)$ is insonified indirectly by backscattered speckle, rather than directly by the transmit pulse (orange segment), ensuring that the point is insonified in transmission mode from below rather than reflection mode from above.      }      
	\label{fig2}
\end{figure} 

The final correspondence between US DPC and optical DIC relates to imaging. In optical DIC, the numerical aperture (NA) of the microscope objective defines the spatial resolution with which sample points are imaged (or, more correctly, the phase differences between \emph{pairs} of sample points). In US DPC, imaging is performed instead numerically by beamforming the raw RF data received by the transducer array. Because US DPC makes use of plane-wave transmits with differing steering angles (at least two), it is naturally compatible with the standard beamforming method of coherent plane-wave compounding \cite{montaldo2009} with no modifications whatsoever. But standard beamforming is designed to perform pulse-echo sonography, which is a single-scattering reflection modality. In particular, standard beamforming assumes that the transmit pulses are directly incident on each reconstructed sample point with the shortest time delay possible (orange segment in Fig. 2b). In our case, US DPC is a transmission modality as opposed to a reflection modality. The insonification arriving at each reconstructed sample point arrives not directly, but rather indirectly by way of backscattering from regions deeper within the sample. For standard beamforming to be tricked into reconstructing indirectly insonified sample points rather than directly insonified sample points, it suffices here to simply add a time delay to the transmit path, as depicted in Fig. 2b (thick blue segments). In other words, it suffices to numerically shift the raw RF data by a delay $T \cos \theta$, where $T$ is a user-defined arbitrary time delay presumed to be reasonably large (more on this below). This delay shift must be performed \emph{prior} to beamforming.

Our algorithm for DPC is summarized below:

\begin{enumerate}
	\item Obtain $RF_n(x,t)$ using a sequence of $N$ transmit plane-wave pulses of tilt angle $\theta_n$ $(n=1...N)$
	\item Numerically delay the raw RF signals: $\widehat{RF}_n(x,t)=RF_n(x,t+T \cos \theta_n)$, where $T$ is user defined 
	\item Perform standard beamforming on each $\widehat{RF}_n(x,t)$, obtaining reconstructed complex images $B_n(x,z)$
	\item Register (recombine) pairs of beamformed images:
	\begin{itemize}
		\item[] $\widehat{B}_n(x,z)=B_n(x-\Delta x/2,z)$
		\item[] $\widehat{B}_{n+m}(x,z)=B_{n+m}(x+\Delta x/2,z)$
	    \item[] where $\Delta x =\frac{cT}{2}(\tan \theta_n-\tan \theta_{n+m})$
	\end{itemize}
    \item Create DPC image from each beamform pair: $I_n(x,z)=\arg \left(\widehat{B}_n(x,z) \widehat{B}_{n+m}^{*}(x,z)\right)$
\end{enumerate}
Note that step 4 was derived from Eq. \ref{DPC1}, and step 5 evaluates the local phase difference between beamform pairs. Note also that standard pulse-echo imaging involves steps 1 and 3 only (without step 2). 

\begin{figure}[tbp]
	\includegraphics[width =2.8in]{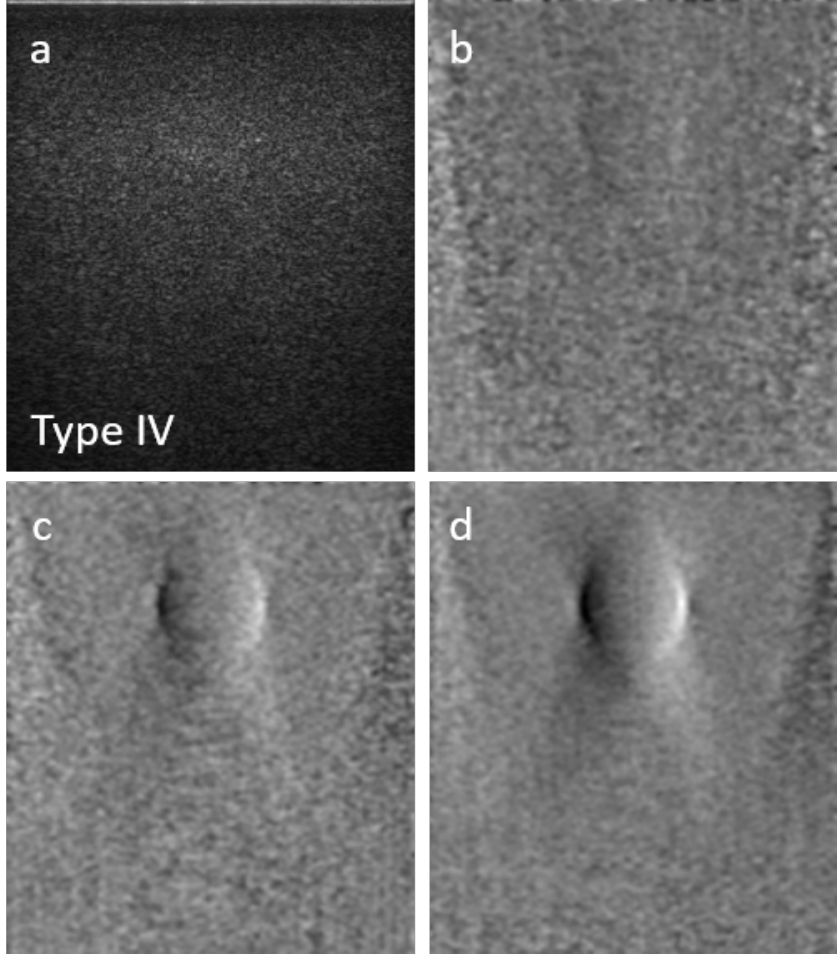}
	\caption{Images of spherical SoS inclusion in scattering phantom. (a) Standard pulse-echo intensity image. DPC images (Gaussian filtered) from (b) single transmit pair, and from multiple transmit pairs with (c) angular and (d) added delay compounding.   }      
	\label{fig3}
\end{figure}   

\begin{figure}[tbp]
	\includegraphics[width =3.3in]{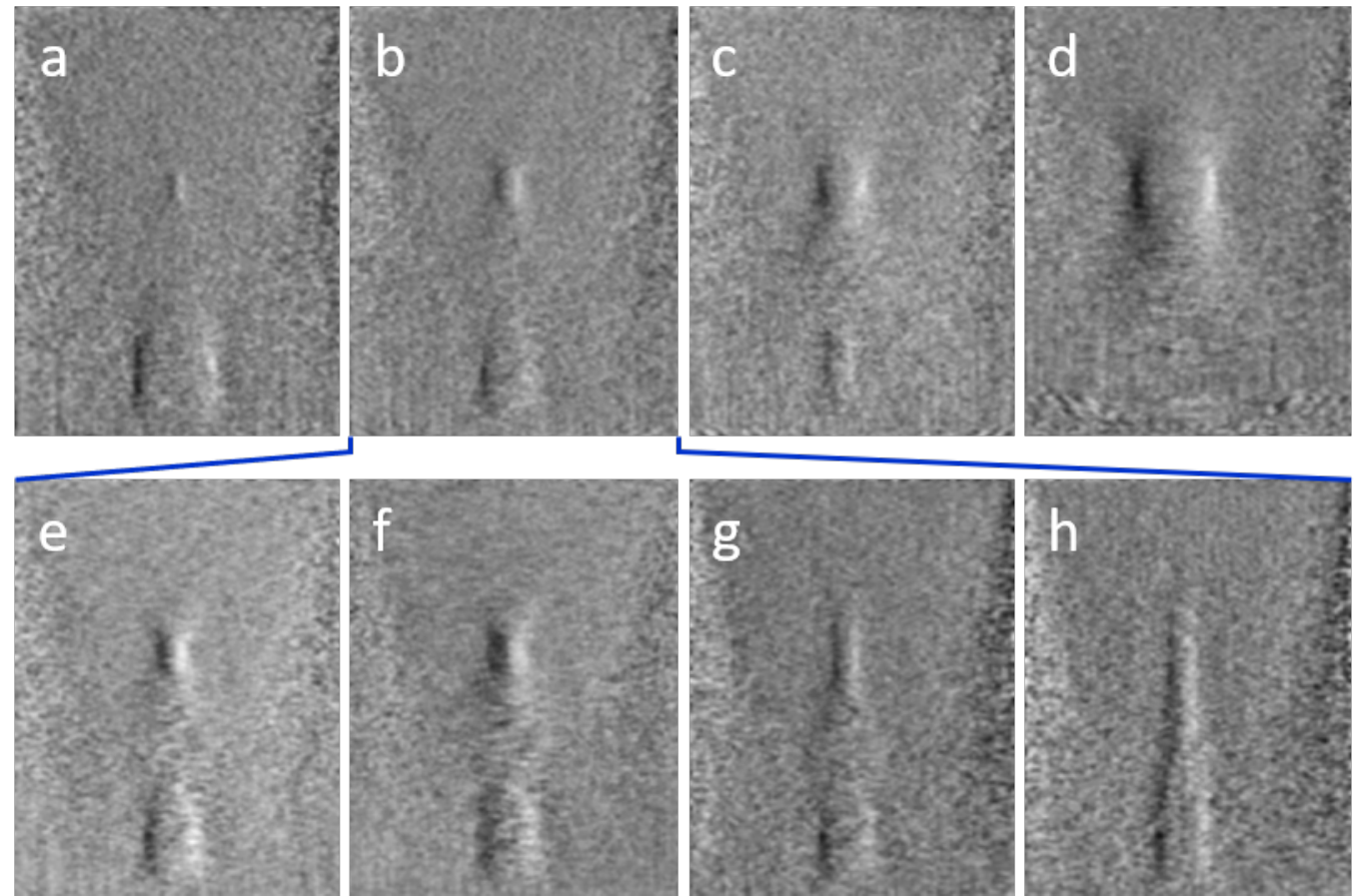}
	\caption{Images of cylindrical SoS inclusions. (a-d) $m=1$, NA $=0.6$. Shallower cylinders increase in diameter (2.5 mm, 4.1 mm, 6.5 mm, 10.4 mm); deeper cylinders decrease in diameter (vice versa). (e,f) Same as (b) but with $m=2$ and $m=4$. (g,h) Same as (b) but with NA $=0.3$ and NA $=0.15$. }      
	\label{fig4}
\end{figure} 

We experimentally demonstrate the performance of our DPC algorithm using the CIRS Model 049 phantom. This phantom contains four types of spherical elasticity inclusions 10 mm in diameter and 15 mm below the phantom surface. The inclusions feature differing SoS from the background SoS (Types I-IV: $1530 \pm 10$ m/s, $1533 \pm 10$ m/s, $1552 \pm 10$ m/s, $1572 \pm 7$ m/s -- according to manufacturer specifications). We insonified the phantom with a sequence of $13$ plane wave transmit pulses with steering angles $\theta_n$ ranging from $-0.15$ to $0.15$ radians, obtaining a sequence of receive signals $RF_n(x,t)$. Figure 3a shows a B-mode image of a Type IV inclusion obtained by standard beamforming with coherent plane-wave compounding \cite{montaldo2009}. The inclusion is barely visible, since it presents little change in attenuation or scattering relative to the background. We then applied our DPC algorithm to the same set of receive signals, with $T$ arbitrarily set to $800$ sampling periods (corresponding to an added echo distance of 58 mm), and with inter-angle separation index $m=1$. Figure 3b shows a resulting DPC image $I_{n=7}(x,z)$ using only a single pair of transmit angles. Figure 3c shows the DPC image $I=\langle I_n \rangle_{n}$ after angular compounding over multiple pairs (12 in total). Figure 3d shows the effect of additional delay compounding $I=\langle I_n \rangle_{n,T}$ with $T=600,800,1000,1200$, involving a $4\times$ increase in the number of beamforming steps. Manifestly, compounding leads to an improvement in SNR, allowing the inclusion to become more readily apparent. We note that Figs. 3c-d bear close resemblance to optical DIC images, except that they are oriented here in the $(x,z)$ plane rather than the $(x,y)$ plane.

Somewhat more involved is the CIRS Model 049A phantom, which contains stepped cylinder inclusions of differing diameters at two deeper depths (30 and 60 mm below surface). DPC images of Type IV inclusions ($T=600$, no delay compounding) are shown in Fig. 4 for various imaging conditions. Figures 4b,e,f show the effect of increasing $m$, which leads to a larger angular separation between transmit pairs and hence to a larger shear separation $\Delta x$ between image pairs. The result is a moderate increase in contrast, though at the cost of degraded transverse resolution. Figures 4b,g,h show the effect of decreasing NA when beamforming. As expected, lowering the NA leads to a rapid elongation of the receive PSFs, significantly undermining axial resolution and highlighting the importance of NA.

\begin{figure}[tbp]
	\includegraphics[width =3.3in]{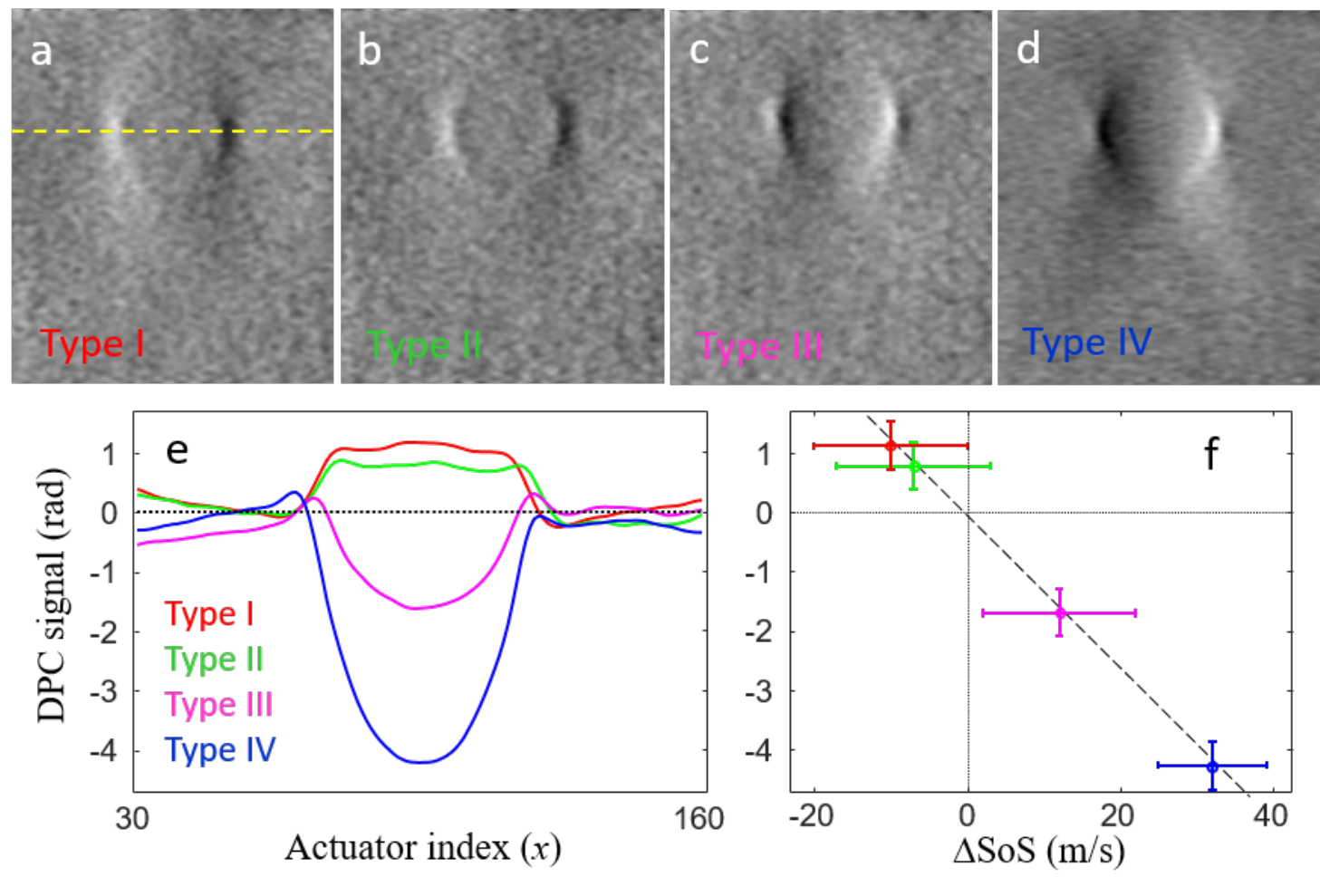}
	\caption{(a-d) DPC images of different types of SoS inclusions. (e) Transverse integration of (smoothed) DPC images across inclusion equators (yellow dashed), showing sign and magnitude of phase shift relative to background. (f) Maximum phase-shift excursion is closely proportional to the difference between inclusion and background SoS, or $\Delta$SoS.     }      
	\label{fig5}
\end{figure} 

Finally, we demonstrate the capacity of DPC to discriminate between different SoS values. Figures 5a-d show DPC images (angular and delay compounded) of the four types of inclusion in the CIRS Model 049 phantom. As expected, the contrast becomes inverted depending on whether the inclusion SoS is greater or smaller than the background. These images reveal the local transverse gradient of the SoS throughout the phantom. This gradient may be integrated along the transverse direction (yellow dashed in Fig. 5a) to provide an estimated map of the SoS itself, to within an integration constant assumed here to be the background SoS (Fig. 5e). Remarkably, the maximum SoS excursion from background at the center of the inclusions appears to closely obey a linear scaling law as a function of SoS (Fig. 5f), suggesting that DPC may ultimately be useful in providing quantitative estimates of SoS, at least for simple sample geometries.

In summary, we have demonstrated a method to perform US phase imaging. Although based on transmission imaging the method works in a reflection configuration, and makes use of standard beamforming based on plane-wave compounding with no additional data requirements and very little additional computation. As such, it can be combined with conventional B-mode imaging while providing complementary phase (or SoS) contrast in real time and essentially for free, enabling sample structures to be revealed that would otherwise be invisible. However, our results come with caveats. The phantoms we utilized were relatively simple, with homogeneous background SoS that was assumed to be known a priori. It remains to be seen how well this technique performs with more complex samples. Moreover, in its current form US DPC only reveals transverse phase gradients, and is not well suited for imaging layered SoS variations such as often encountered in practice. Nevertheless, these preliminary results provide a basis for a promising new modality in US imaging, and for imaging in scattering media using the memory effect \cite{bertolotti2012,katz2014}.

This work was funded by NIH R21GM134216. We thank Thomas Bifano and the BU Photonics Center for making a Verasonics machine available to us, and Michael Jaeger for insightful discussions.


%



\begin{thebibliography}{23}%
	\makeatletter
	\providecommand \@ifxundefined [1]{%
		\@ifx{#1\undefined}
	}%
	\providecommand \@ifnum [1]{%
		\ifnum #1\expandafter \@firstoftwo
		\else \expandafter \@secondoftwo
		\fi
	}%
	\providecommand \@ifx [1]{%
		\ifx #1\expandafter \@firstoftwo
		\else \expandafter \@secondoftwo
		\fi
	}%
	\providecommand \natexlab [1]{#1}%
	\providecommand \enquote  [1]{``#1''}%
	\providecommand \bibnamefont  [1]{#1}%
	\providecommand \bibfnamefont [1]{#1}%
	\providecommand \citenamefont [1]{#1}%
	\providecommand \href@noop [0]{\@secondoftwo}%
	\providecommand \href [0]{\begingroup \@sanitize@url \@href}%
	\providecommand \@href[1]{\@@startlink{#1}\@@href}%
	\providecommand \@@href[1]{\endgroup#1\@@endlink}%
	\providecommand \@sanitize@url [0]{\catcode `\\12\catcode `\$12\catcode
		`\&12\catcode `\#12\catcode `\^12\catcode `\_12\catcode `\%12\relax}%
	\providecommand \@@startlink[1]{}%
	\providecommand \@@endlink[0]{}%
	\providecommand \url  [0]{\begingroup\@sanitize@url \@url }%
	\providecommand \@url [1]{\endgroup\@href {#1}{\urlprefix }}%
	\providecommand \urlprefix  [0]{URL }%
	\providecommand \Eprint [0]{\href }%
	\providecommand \doibase [0]{https://doi.org/}%
	\providecommand \selectlanguage [0]{\@gobble}%
	\providecommand \bibinfo  [0]{\@secondoftwo}%
	\providecommand \bibfield  [0]{\@secondoftwo}%
	\providecommand \translation [1]{[#1]}%
	\providecommand \BibitemOpen [0]{}%
	\providecommand \bibitemStop [0]{}%
	\providecommand \bibitemNoStop [0]{.\EOS\space}%
	\providecommand \EOS [0]{\spacefactor3000\relax}%
	\providecommand \BibitemShut  [1]{\csname bibitem#1\endcsname}%
	\let\auto@bib@innerbib\@empty
	\bibitem [{\citenamefont {Anderson}\ \emph {et~al.}(2000)\citenamefont
		{Anderson}, \citenamefont {McKeag},\ and\ \citenamefont
		{Trahey}}]{anderson2000}%
	\BibitemOpen
	\bibfield  {author} {\bibinfo {author} {\bibfnamefont {M.}~\bibnamefont
			{Anderson}}, \bibinfo {author} {\bibfnamefont {M.}~\bibnamefont {McKeag}},\
		and\ \bibinfo {author} {\bibfnamefont {G.}~\bibnamefont {Trahey}},\
	}\href@noop {} {\bibfield  {journal} {\bibinfo  {journal} {Journal of the
				Acoustical Society of America}\ }\textbf {\bibinfo {volume} {107}},\ \bibinfo
		{pages} {3540} (\bibinfo {year} {2000})}\BibitemShut {NoStop}%
	\bibitem [{\citenamefont {Flax}\ and\ \citenamefont
		{O'Donnell}(1988)}]{flax1988}%
	\BibitemOpen
	\bibfield  {author} {\bibinfo {author} {\bibfnamefont {S.}~\bibnamefont
			{Flax}}\ and\ \bibinfo {author} {\bibfnamefont {M.}~\bibnamefont
			{O'Donnell}},\ }\href@noop {} {\bibfield  {journal} {\bibinfo  {journal}
			{IEEE transactions on ultrasonics, ferroelectrics, and frequency control}\
		}\textbf {\bibinfo {volume} {35}},\ \bibinfo {pages} {758} (\bibinfo {year}
		{1988})}\BibitemShut {NoStop}%
	\bibitem [{\citenamefont {Nock}\ \emph {et~al.}(1989)\citenamefont {Nock},
		\citenamefont {Trahey},\ and\ \citenamefont {Smith}}]{nock1989}%
	\BibitemOpen
	\bibfield  {author} {\bibinfo {author} {\bibfnamefont {L.}~\bibnamefont
			{Nock}}, \bibinfo {author} {\bibfnamefont {G.~E.}\ \bibnamefont {Trahey}},\
		and\ \bibinfo {author} {\bibfnamefont {S.~W.}\ \bibnamefont {Smith}},\
	}\href@noop {} {\bibfield  {journal} {\bibinfo  {journal} {Journal of the
				Acoustical Society of America}\ }\textbf {\bibinfo {volume} {85}},\ \bibinfo
		{pages} {1819} (\bibinfo {year} {1989})}\BibitemShut {NoStop}%
	\bibitem [{\citenamefont {Liu}\ and\ \citenamefont {Waag}(1994)}]{liu1994}%
	\BibitemOpen
	\bibfield  {author} {\bibinfo {author} {\bibfnamefont {D.-L.}\ \bibnamefont
			{Liu}}\ and\ \bibinfo {author} {\bibfnamefont {R.~C.}\ \bibnamefont {Waag}},\
	}\href@noop {} {\bibfield  {journal} {\bibinfo  {journal} {Journal of the
				Acoustical Society of America}\ }\textbf {\bibinfo {volume} {96}},\ \bibinfo
		{pages} {649} (\bibinfo {year} {1994})}\BibitemShut {NoStop}%
	\bibitem [{\citenamefont {Lin}\ \emph {et~al.}(1987)\citenamefont {Lin},
		\citenamefont {Ophir},\ and\ \citenamefont {Potter}}]{lin1987}%
	\BibitemOpen
	\bibfield  {author} {\bibinfo {author} {\bibfnamefont {T.}~\bibnamefont
			{Lin}}, \bibinfo {author} {\bibfnamefont {J.}~\bibnamefont {Ophir}},\ and\
		\bibinfo {author} {\bibfnamefont {G.}~\bibnamefont {Potter}},\ }\href@noop {}
	{\bibfield  {journal} {\bibinfo  {journal} {Ultrasonic Imaging}\ }\textbf
		{\bibinfo {volume} {9}},\ \bibinfo {pages} {29} (\bibinfo {year}
		{1987})}\BibitemShut {NoStop}%
	\bibitem [{\citenamefont {Li}\ \emph {et~al.}(2009)\citenamefont {Li},
		\citenamefont {Duric}, \citenamefont {Littrup},\ and\ \citenamefont
		{Huang}}]{li2009}%
	\BibitemOpen
	\bibfield  {author} {\bibinfo {author} {\bibfnamefont {C.}~\bibnamefont
			{Li}}, \bibinfo {author} {\bibfnamefont {N.}~\bibnamefont {Duric}}, \bibinfo
		{author} {\bibfnamefont {P.}~\bibnamefont {Littrup}},\ and\ \bibinfo {author}
		{\bibfnamefont {L.}~\bibnamefont {Huang}},\ }\href@noop {} {\bibfield
		{journal} {\bibinfo  {journal} {Ultrasound in Medicine \& Biology}\ }\textbf
		{\bibinfo {volume} {35}},\ \bibinfo {pages} {1615} (\bibinfo {year}
		{2009})}\BibitemShut {NoStop}%
	\bibitem [{\citenamefont {Glozman}\ and\ \citenamefont
		{Azhari}(2010)}]{glozman2010}%
	\BibitemOpen
	\bibfield  {author} {\bibinfo {author} {\bibfnamefont {T.}~\bibnamefont
			{Glozman}}\ and\ \bibinfo {author} {\bibfnamefont {H.}~\bibnamefont
			{Azhari}},\ }\href@noop {} {\bibfield  {journal} {\bibinfo  {journal}
			{Journal of Ultrasound in Medicine}\ }\textbf {\bibinfo {volume} {29}},\
		\bibinfo {pages} {387} (\bibinfo {year} {2010})}\BibitemShut {NoStop}%
	\bibitem [{\citenamefont {Duric}\ \emph {et~al.}(2013)\citenamefont {Duric},
		\citenamefont {Boyd}, \citenamefont {Littrup}, \citenamefont {Sak},
		\citenamefont {Myc}, \citenamefont {Li}, \citenamefont {West}, \citenamefont
		{Minkin}, \citenamefont {Martin}, \citenamefont {Yaffe} \emph
		{et~al.}}]{duric2013}%
	\BibitemOpen
	\bibfield  {author} {\bibinfo {author} {\bibfnamefont {N.}~\bibnamefont
			{Duric}}, \bibinfo {author} {\bibfnamefont {N.}~\bibnamefont {Boyd}},
		\bibinfo {author} {\bibfnamefont {P.}~\bibnamefont {Littrup}}, \bibinfo
		{author} {\bibfnamefont {M.}~\bibnamefont {Sak}}, \bibinfo {author}
		{\bibfnamefont {L.}~\bibnamefont {Myc}}, \bibinfo {author} {\bibfnamefont
			{C.}~\bibnamefont {Li}}, \bibinfo {author} {\bibfnamefont {E.}~\bibnamefont
			{West}}, \bibinfo {author} {\bibfnamefont {S.}~\bibnamefont {Minkin}},
		\bibinfo {author} {\bibfnamefont {L.}~\bibnamefont {Martin}}, \bibinfo
		{author} {\bibfnamefont {M.}~\bibnamefont {Yaffe}}, \emph {et~al.},\
	}\href@noop {} {\bibfield  {journal} {\bibinfo  {journal} {Medical Physics}\
		}\textbf {\bibinfo {volume} {40}},\ \bibinfo {pages} {013501} (\bibinfo
		{year} {2013})}\BibitemShut {NoStop}%
	\bibitem [{\citenamefont {Nebeker}\ and\ \citenamefont
		{Nelson}(2012)}]{nebeker2012}%
	\BibitemOpen
	\bibfield  {author} {\bibinfo {author} {\bibfnamefont {J.}~\bibnamefont
			{Nebeker}}\ and\ \bibinfo {author} {\bibfnamefont {T.~R.}\ \bibnamefont
			{Nelson}},\ }\href@noop {} {\bibfield  {journal} {\bibinfo  {journal}
			{Journal of Ultrasound in Medicine}\ }\textbf {\bibinfo {volume} {31}},\
		\bibinfo {pages} {1389} (\bibinfo {year} {2012})}\BibitemShut {NoStop}%
	\bibitem [{\citenamefont {Robinson}\ \emph {et~al.}(1991)\citenamefont
		{Robinson}, \citenamefont {Ophir}, \citenamefont {Wilson},\ and\
		\citenamefont {Chen}}]{robinson1991}%
	\BibitemOpen
	\bibfield  {author} {\bibinfo {author} {\bibfnamefont {D.}~\bibnamefont
			{Robinson}}, \bibinfo {author} {\bibfnamefont {J.}~\bibnamefont {Ophir}},
		\bibinfo {author} {\bibfnamefont {L.}~\bibnamefont {Wilson}},\ and\ \bibinfo
		{author} {\bibfnamefont {C.}~\bibnamefont {Chen}},\ }\href@noop {} {\bibfield
		{journal} {\bibinfo  {journal} {Ultrasound in Medicine \& Biology}\ }\textbf
		{\bibinfo {volume} {17}},\ \bibinfo {pages} {633 } (\bibinfo {year}
		{1991})}\BibitemShut {NoStop}%
	\bibitem [{\citenamefont {Imbault}\ \emph {et~al.}(2017)\citenamefont
		{Imbault}, \citenamefont {Faccinetto}, \citenamefont {Osmanski},
		\citenamefont {Tissier}, \citenamefont {Deffieux}, \citenamefont {Gennisson},
		\citenamefont {Vilgrain},\ and\ \citenamefont {Tanter}}]{imbault2017}%
	\BibitemOpen
	\bibfield  {author} {\bibinfo {author} {\bibfnamefont {M.}~\bibnamefont
			{Imbault}}, \bibinfo {author} {\bibfnamefont {A.}~\bibnamefont {Faccinetto}},
		\bibinfo {author} {\bibfnamefont {B.-F.}\ \bibnamefont {Osmanski}}, \bibinfo
		{author} {\bibfnamefont {A.}~\bibnamefont {Tissier}}, \bibinfo {author}
		{\bibfnamefont {T.}~\bibnamefont {Deffieux}}, \bibinfo {author}
		{\bibfnamefont {J.-L.}\ \bibnamefont {Gennisson}}, \bibinfo {author}
		{\bibfnamefont {V.}~\bibnamefont {Vilgrain}},\ and\ \bibinfo {author}
		{\bibfnamefont {M.}~\bibnamefont {Tanter}},\ }\href@noop {} {\bibfield
		{journal} {\bibinfo  {journal} {Physics in Medicine \& Biology}\ }\textbf
		{\bibinfo {volume} {62}},\ \bibinfo {pages} {3582} (\bibinfo {year}
		{2017})}\BibitemShut {NoStop}%
	\bibitem [{\citenamefont {Lambert}\ \emph {et~al.}(2020)\citenamefont
		{Lambert}, \citenamefont {Cobus}, \citenamefont {Couade}, \citenamefont
		{Fink},\ and\ \citenamefont {Aubry}}]{lambert2020}%
	\BibitemOpen
	\bibfield  {author} {\bibinfo {author} {\bibfnamefont {W.}~\bibnamefont
			{Lambert}}, \bibinfo {author} {\bibfnamefont {L.~A.}\ \bibnamefont {Cobus}},
		\bibinfo {author} {\bibfnamefont {M.}~\bibnamefont {Couade}}, \bibinfo
		{author} {\bibfnamefont {M.}~\bibnamefont {Fink}},\ and\ \bibinfo {author}
		{\bibfnamefont {A.}~\bibnamefont {Aubry}},\ }\href
	{https://doi.org/10.1103/PhysRevX.10.021048} {\bibfield  {journal} {\bibinfo
			{journal} {Phys. Rev. X}\ }\textbf {\bibinfo {volume} {10}},\ \bibinfo
		{pages} {021048} (\bibinfo {year} {2020})}\BibitemShut {NoStop}%
	\bibitem [{\citenamefont {Kr\"{u}cker}\ \emph {et~al.}(2004)\citenamefont
		{Kr\"{u}cker}, \citenamefont {Fowlkes},\ and\ \citenamefont
		{Carson}}]{krucker2004}%
	\BibitemOpen
	\bibfield  {author} {\bibinfo {author} {\bibfnamefont {J.}~\bibnamefont
			{Kr\"{u}cker}}, \bibinfo {author} {\bibfnamefont {J.~B.}\ \bibnamefont
			{Fowlkes}},\ and\ \bibinfo {author} {\bibfnamefont {P.~L.}\ \bibnamefont
			{Carson}},\ }\href@noop {} {\bibfield  {journal} {\bibinfo  {journal} {IEEE
				Transactions on Ultrasonics, Ferroelectrics, and Frequency Control}\ }\textbf
		{\bibinfo {volume} {51}},\ \bibinfo {pages} {1095} (\bibinfo {year}
		{2004})}\BibitemShut {NoStop}%
	\bibitem [{\citenamefont {Stähli}\ \emph {et~al.}(2020)\citenamefont
		{Stähli}, \citenamefont {Kuriakose}, \citenamefont {Frenz},\ and\
		\citenamefont {Jaeger}}]{stahli2020}%
	\BibitemOpen
	\bibfield  {author} {\bibinfo {author} {\bibfnamefont {P.}~\bibnamefont
			{Stähli}}, \bibinfo {author} {\bibfnamefont {M.}~\bibnamefont {Kuriakose}},
		\bibinfo {author} {\bibfnamefont {M.}~\bibnamefont {Frenz}},\ and\ \bibinfo
		{author} {\bibfnamefont {M.}~\bibnamefont {Jaeger}},\ }\href
	{https://doi.org/https://doi.org/10.1016/j.ultras.2020.106168} {\bibfield
		{journal} {\bibinfo  {journal} {Ultrasonics}\ }\textbf {\bibinfo {volume}
			{108}},\ \bibinfo {pages} {106168} (\bibinfo {year} {2020})}\BibitemShut
	{NoStop}%
	\bibitem [{\citenamefont {Pluta}(1994)}]{pluta1994}%
	\BibitemOpen
	\bibfield  {author} {\bibinfo {author} {\bibfnamefont {M.}~\bibnamefont
			{Pluta}},\ }\href {https://doi.org/10.1117/12.171873} {\bibfield  {journal}
		{\bibinfo  {journal} {Proc. SPIE}\ }\textbf {\bibinfo {volume} {1846}},\
		\bibinfo {pages} {10 } (\bibinfo {year} {1994})}\BibitemShut {NoStop}%
	\bibitem [{\citenamefont {Mallart}\ and\ \citenamefont
		{Fink}(1991)}]{mallart1991}%
	\BibitemOpen
	\bibfield  {author} {\bibinfo {author} {\bibfnamefont {R.}~\bibnamefont
			{Mallart}}\ and\ \bibinfo {author} {\bibfnamefont {M.}~\bibnamefont {Fink}},\
	}\href {https://doi.org/10.1121/1.401867} {\bibfield  {journal} {\bibinfo
			{journal} {Journal of the Acoustical Society of America}\ }\textbf {\bibinfo
			{volume} {90}},\ \bibinfo {pages} {2718} (\bibinfo {year}
		{1991})}\BibitemShut {NoStop}%
	\bibitem [{\citenamefont {Feng}\ \emph {et~al.}(1988)\citenamefont {Feng},
		\citenamefont {Kane}, \citenamefont {Lee},\ and\ \citenamefont
		{Stone}}]{Feng1988}%
	\BibitemOpen
	\bibfield  {author} {\bibinfo {author} {\bibfnamefont {S.}~\bibnamefont
			{Feng}}, \bibinfo {author} {\bibfnamefont {C.}~\bibnamefont {Kane}}, \bibinfo
		{author} {\bibfnamefont {P.~A.}\ \bibnamefont {Lee}},\ and\ \bibinfo {author}
		{\bibfnamefont {A.~D.}\ \bibnamefont {Stone}},\ }\href@noop {} {\bibfield
		{journal} {\bibinfo  {journal} {Phys. Rev. Lett.}\ }\textbf {\bibinfo
			{volume} {61}},\ \bibinfo {pages} {834} (\bibinfo {year} {1988})}\BibitemShut
	{NoStop}%
	\bibitem [{\citenamefont {Berkovits}\ \emph {et~al.}(1989)\citenamefont
		{Berkovits}, \citenamefont {Kaveh},\ and\ \citenamefont
		{Feng}}]{berkovitz1989}%
	\BibitemOpen
	\bibfield  {author} {\bibinfo {author} {\bibfnamefont {R.}~\bibnamefont
			{Berkovits}}, \bibinfo {author} {\bibfnamefont {M.}~\bibnamefont {Kaveh}},\
		and\ \bibinfo {author} {\bibfnamefont {S.}~\bibnamefont {Feng}},\ }\href
	{https://doi.org/10.1103/PhysRevB.40.737} {\bibfield  {journal} {\bibinfo
			{journal} {Phys. Rev. B}\ }\textbf {\bibinfo {volume} {40}},\ \bibinfo
		{pages} {737} (\bibinfo {year} {1989})}\BibitemShut {NoStop}%
	\bibitem [{\citenamefont {Freund}(1990)}]{freund1990}%
	\BibitemOpen
	\bibfield  {author} {\bibinfo {author} {\bibfnamefont {I.}~\bibnamefont
			{Freund}},\ }\href@noop {} {\bibfield  {journal} {\bibinfo  {journal}
			{Physica A: Statistical Mechanics and its Applications}\ }\textbf {\bibinfo
			{volume} {168}},\ \bibinfo {pages} {49 } (\bibinfo {year}
		{1990})}\BibitemShut {NoStop}%
	\bibitem [{\citenamefont {Kadobianskyi}\ \emph {et~al.}(2018)\citenamefont
		{Kadobianskyi}, \citenamefont {Papadopoulos}, \citenamefont {Chaigne},
		\citenamefont {Horstmeyer},\ and\ \citenamefont
		{Judkewitz}}]{Kadobianskyi2018}%
	\BibitemOpen
	\bibfield  {author} {\bibinfo {author} {\bibfnamefont {M.}~\bibnamefont
			{Kadobianskyi}}, \bibinfo {author} {\bibfnamefont {I.~N.}\ \bibnamefont
			{Papadopoulos}}, \bibinfo {author} {\bibfnamefont {T.}~\bibnamefont
			{Chaigne}}, \bibinfo {author} {\bibfnamefont {R.}~\bibnamefont
			{Horstmeyer}},\ and\ \bibinfo {author} {\bibfnamefont {B.}~\bibnamefont
			{Judkewitz}},\ }\href@noop {} {\bibfield  {journal} {\bibinfo  {journal}
			{Optica}\ }\textbf {\bibinfo {volume} {5}},\ \bibinfo {pages} {389} (\bibinfo
		{year} {2018})}\BibitemShut {NoStop}%
	\bibitem [{\citenamefont {Montaldo}\ \emph {et~al.}(2009)\citenamefont
		{Montaldo}, \citenamefont {Tanter}, \citenamefont {Bercoff}, \citenamefont
		{Benech},\ and\ \citenamefont {Fink}}]{montaldo2009}%
	\BibitemOpen
	\bibfield  {author} {\bibinfo {author} {\bibfnamefont {G.}~\bibnamefont
			{Montaldo}}, \bibinfo {author} {\bibfnamefont {M.}~\bibnamefont {Tanter}},
		\bibinfo {author} {\bibfnamefont {J.}~\bibnamefont {Bercoff}}, \bibinfo
		{author} {\bibfnamefont {N.}~\bibnamefont {Benech}},\ and\ \bibinfo {author}
		{\bibfnamefont {M.}~\bibnamefont {Fink}},\ }\href@noop {} {\bibfield
		{journal} {\bibinfo  {journal} {IEEE Transactions on Ultrasonics,
				Ferroelectrics, and Frequency Control}\ }\textbf {\bibinfo {volume} {56}},\
		\bibinfo {pages} {489} (\bibinfo {year} {2009})}\BibitemShut {NoStop}%
	\bibitem [{\citenamefont {Bertolotti}\ \emph {et~al.}(2012)\citenamefont
		{Bertolotti}, \citenamefont {Van~Putten}, \citenamefont {Blum}, \citenamefont
		{Lagendijk}, \citenamefont {Vos},\ and\ \citenamefont
		{Mosk}}]{bertolotti2012}%
	\BibitemOpen
	\bibfield  {author} {\bibinfo {author} {\bibfnamefont {J.}~\bibnamefont
			{Bertolotti}}, \bibinfo {author} {\bibfnamefont {E.~G.}\ \bibnamefont
			{Van~Putten}}, \bibinfo {author} {\bibfnamefont {C.}~\bibnamefont {Blum}},
		\bibinfo {author} {\bibfnamefont {A.}~\bibnamefont {Lagendijk}}, \bibinfo
		{author} {\bibfnamefont {W.~L.}\ \bibnamefont {Vos}},\ and\ \bibinfo {author}
		{\bibfnamefont {A.~P.}\ \bibnamefont {Mosk}},\ }\href@noop {} {\bibfield
		{journal} {\bibinfo  {journal} {Nature}\ }\textbf {\bibinfo {volume} {491}},\
		\bibinfo {pages} {232} (\bibinfo {year} {2012})}\BibitemShut {NoStop}%
	\bibitem [{\citenamefont {Katz}\ \emph {et~al.}(2014)\citenamefont {Katz},
		\citenamefont {Heidmann}, \citenamefont {Fink},\ and\ \citenamefont
		{Gigan}}]{katz2014}%
	\BibitemOpen
	\bibfield  {author} {\bibinfo {author} {\bibfnamefont {O.}~\bibnamefont
			{Katz}}, \bibinfo {author} {\bibfnamefont {P.}~\bibnamefont {Heidmann}},
		\bibinfo {author} {\bibfnamefont {M.}~\bibnamefont {Fink}},\ and\ \bibinfo
		{author} {\bibfnamefont {S.}~\bibnamefont {Gigan}},\ }\href@noop {}
	{\bibfield  {journal} {\bibinfo  {journal} {Nature Photonics}\ }\textbf
		{\bibinfo {volume} {8}},\ \bibinfo {pages} {784} (\bibinfo {year}
		{2014})}\BibitemShut {NoStop}%
\end{thebibliography}
\end{document}